\documentclass[12pt]{article}
\usepackage{amssymb, amsmath}
\usepackage{epsfig}
\usepackage{array}
\makeatletter
\def\alt{\mathrel{\mathpalette\gl@align<}}
\def\agt{\mathrel{\mathpalette\gl@align>}}
\def\gl@align#1#2{\lower.6ex\vbox{\baselineskip\z@skip\lineskip\z@
    \ialign{$\m@th#1\hfil##\hfil$\crcr#2\crcr\sim\crcr}}} \makeatother

\def \gtsim    {\relax\ifmmode{\mathrel{\mathpalette\oversim >}}
  \else{$\mathrel{\mathpalette\oversim >}$}\fi}
\def \ltsim    {\relax\ifmmode{\mathrel{\mathpalette\oversim <}}
  \else{$\mathrel{\mathpalette\oversim <}$}\fi}
\def\oversim#1#2{\lower4pt\vbox{\baselineskip0pt \lineskip1.5pt
    \ialign{$\mathsurround=0pt#1\hfil##\hfil$\crcr#2\crcr\sim\crcr}}}

\newcommand{\mett}{\mbox{$E\!\!\!\!/_{T}$}}

\def \Et {\mbox{$E_T$}}

\def\goes{\mbox{$\rightarrow$}}

\def\ttbar{\mbox{$t{\bar t}$}}

\newcommand{\squark}{\mbox{$\widetilde{q}$}}
\newcommand{\squarkr}{\mbox{$\widetilde{q}_{R}$}}
\newcommand{\gluino}{\mbox{$\widetilde{g}$}}
\newcommand{\bbar}{\mbox{$\bar{b}$}}
\newcommand{\tbar}{\mbox{$\bar{t}$}}
\newcommand{\tops}{\mbox{$\widetilde{t}_{1}$}}
\newcommand{\sbottom}{\mbox{$\widetilde{b}_{1}$}}
\newcommand{\ett}{\Et}

\newcommand{\NONE}{\mbox{$\widetilde{\chi}_1^0$}}
\newcommand{\NTWO}{\mbox{$\widetilde{\chi}_2^0$}}
\newcommand{\CONE}{\mbox{$\widetilde{\chi}_1^{\pm}$}}
\newcommand{\CONEP}{\mbox{$\widetilde{\chi}_1^{+}$}}
\newcommand{\CONEM}{\mbox{$\widetilde{\chi}_1^{-}$}}
\newcommand{\stau}{\mbox{$\widetilde{\tau}$}}
\newcommand{\none}{\NONE}
\newcommand{\ntwo}{\NTWO}
\newcommand{\cone}{\CONE}

\newcommand{\tanbeta}{\mbox{$\tan\beta$}}
\newcommand{\nosls}{\mbox{$N_{OS-LS}$}}
\newcommand{\mtt}{\mbox{$M_{\tau\tau}^{\mathrm{peak}}$}}

\newcommand{\Plot}[3]{
\begin{figure}[!htb]
\centering{\epsfig{file=#1,width=0.4\linewidth}}
\caption{#2}
\label{#3}
\end{figure}}

\newcommand{\TwoPlot}[4]{
\begin{figure}[!htb]
\centering{\epsfig{file=#1,width=0.4\linewidth}}
\centering{\epsfig{file=#2,width=0.4\linewidth}}
\caption{#3}
\label{#4}
\end{figure}}

\newcommand{\ThreePlot}[5]{
\begin{figure}[!htb]
\centering{\epsfig{file=#1,width=0.4\linewidth}}
\centering{\epsfig{file=#2,width=0.4\linewidth}}
\centering{\epsfig{file=#3,width=0.4\linewidth}}
\caption{#4}
\label{#5}
\end{figure}}

\newcommand{\FourPlot}[6]{
\begin{figure}[!htb]
\centering{\epsfig{file=#1,width=0.4\linewidth}}
\centering{\epsfig{file=#2,width=0.4\linewidth}}
\centering{\epsfig{file=#3,width=0.4\linewidth}}
\centering{\epsfig{file=#4,width=0.4\linewidth}}
\caption{#5}
\label{#6}
\end{figure}}

\def\mg{\mbox{$M_{\tilde{g}}$}}
\def\dm{\mbox{$\Delta{M}$}}

\newcommand{\myclearpage}{\clearpage}

\newcommand{\mysect}[1]{}
\newcommand{\mytoc}{}

\begin{document}

\begin{flushright}
MIFP-0621
\end{flushright}
\vspace*{2cm}


\begin{center}
{\baselineskip 25pt

\large{\bf Measurement of the \stau\ - \none\ Mass Difference and \mg\ in the Co-Annihilation Region at the LHC  }\\
}

\vspace{1cm}

{\large Richard Arnowitt,$^1$ Adam Aurisano,$^1$ Bhaskar Dutta,$^1$ Teruki Kamon,$^1$
\\ Nikolay Kolev,$^2$ David Toback$^1$, Paul Simeon,$^1$ and Peter Wagner$^1$} \vspace{.5cm}

{ \it $^1$ Department of Physics, Texas A\&M University, College Station, TX 77843-4242, USA\\
\it $^2$Department of Physics, University of Regina, Regina, SK S4S 0A2, Canada }
\vspace{.5cm}

\vspace{1.5cm} {\bf Abstract}
\end{center}

We study the prospects for the measurement of the \stau\ - \none\ mass difference (\dm) and the \gluino\ mass (\mg) in the supersymmetric co-annihilation region at the LHC using tau leptons. Recent WMAP measurements of the amount of  cold dark matter and previous accelerator experiments indicate that the  coannihilation region of mSUGRA is characterized by a small \dm\ (5-15~GeV).  Focusing on taus from \mbox{$\ntwo\goes\tau\stau\goes\tau\tau\none$} decays in \gluino\ and \squark\ production, we consider inclusive \mbox{3$\tau$+jet+\mett} production, with two $\tau$'s above a high \Et\ threshold and a third $\tau$ above a lower threshold.  Two observables, the number of opposite-signed $\tau$ pairs minus the number of like-signed $\tau$ pairs and the peak of the ditau invariant mass distribution, allow for the simultaneous determination of \dm\ and \mg\ for \dm\ $>$5 GeV.  For example, for  \dm\ = 9~GeV and \mg\ = 850~GeV and  with 30~fb$^{-1}$ of data, we can measure \dm\ to 15\% and \mg\ to 6\%.

\thispagestyle{empty}

\bigskip
\newpage
\addtocounter{page}{-1}

\myclearpage

\mytoc


\newcounter{myctr}


\section{Introduction}
\label{sec:introduction}

\mysect{Intro to SUSY}

Supersymmetry (SUSY), a symmetry between fermions and bosons, allows for the construction of models that link a wide range of physical phenomena. While initially proposed on aesthetic grounds, SUSY also allows for cancellation of the quadratic Higgs divergence and, hence, opens the window for consistent model building up to the grand unification (GUT) or Planck scales.  The extension of supersymmetry to a local gauge theory, supergravity \cite{sugra}, led to the development of GUT models \cite{gut,gut_two} giving a description of physics from the GUT scale down to the electroweak scale, in addition to incorporating the successes of the Standard Model (SM).  LEP data confirmed the validity of the idea of SUSY grand unification.  Additionally,  SUSY models with R-parity conservation automatically give rise to a cold dark matter candidate, the lightest neutralino, \none.  Because these models are consistent to very high energies, they provide descriptions of the early universe that deeply link particle physics and cosmology.  Detailed theoretical calculations \cite{dark_constraint} confirm that GUT models can account for the experimentally observed dark matter abundance \cite{wmap}.

\mysect{Intro to mSUGRA}

There is good reason to believe that SUSY can be discovered at the Large Hadron Collider (LHC) and that it may be possible to determine its parameters with enough precision to distinguish among SUSY models that correctly predict the dark matter content of the universe.  While the ideal is to study the prospects of measuring SUSY in as general a way as possible, we consider mSUGRA \cite{gut} here, a commonly studied model, for concreteness. In mSUGRA, four parameters and one sign determine all the masses and couplings: $m_0$, the universal scalar soft breaking mass at $M_{GUT}$;  $m_{1/2}$, the universal gaugino mass at $M_{GUT}$; $A_{0}$, the universal cubic soft breaking parameter at $M_{GUT}$; \tanbeta = $\langle H_{1} \rangle / \langle H_{2} \rangle$, where $\langle H_{1(2)} \rangle $ is the Higgs vacuum expectation value which gives rise to the up (down) quark masses; and the sign of $\mu$, where $\mu$ is the coefficient of the quadratic term of the superpotential, $W = \mu H_{1} H_{2}$.  Dark matter abundance and accelerator data constrain the parameter space to only three regions \cite{dark_constraint}: (1) the \mbox{\stau - \none} co-annihilation region\footnote{Note that the notation  \stau\ here refers to the lighter of the two supersymmetric tau mass eigenstates.} \cite{coannihilation} where the \none\ is the LSP and \mbox{\dm\ $\equiv$ $M_{\stau}-M_{\none}$}  is small (\dm\ $\sim$ 5-15 GeV),  (2) the focus point region where the \none\ has a large Higgsino component, and (3) the scalar Higgs annihilation funnel region.  A  bulk region is also allowed by dark matter constraints, but it is largely excluded by other constraints.

\mysect{why Co-annihilation?}

A collection of other constraints suggests that, of these regions, the co-annihilation region is particularly important.  They include the following: the light Higgs and lightest chargino mass bounds from LEP \cite{bound}, the \mbox{$b\rightarrow s\gamma$} branching ratio bound \cite{rare_decay}, and the 3.3$\sigma$ deviation from the SM expectation of the anomalous muon magnetic moment from the muon $g-2$ collaboration \cite{muon}.  If the anomalous muon magnetic moment deviation is proven, mostly the co-annihilation region will be allowed. Dark matter production and annihilation rates in the early universe are believed to control the dark matter relic density we observe today; however, co-annihilation can provide an additional mechanism for the reduction of dark matter abundances in the early universe. In particular,  the \stau\ and \none\ can co-annihilate sufficiently into SM particles only if \dm\ is small.  Since, in principle, $M_{\stau}$ and $M_{\none}$ can separately range from 100~GeV into the TeV domain, if this striking smallness of \dm\ were found at the LHC, it would be a strong indirect indication that the \none\ is the astronomical dark matter particle.  We also note that many other SUGRA models have a co-annihilation region (since co-annihilation can occur with non-universal gaugino masses), so investigating this region has greater applicability than just mSUGRA (though, constraints do not necessarily exist in other SUGRA models to select out the co-annihilation region).

\mysect{Previous work}

The co-annihilation region is difficult to analyze due to the production of low energy $\tau$'s from \mbox{$\stau\rightarrow\tau\none$} decays.  Some previous work has been done to determine how to measure masses of SUSY particles in the co-annihilation region. One Monte Carlo study  investigated the co-annihilation region by fitting various final state mass distributions in order to determine the SUSY masses \cite{endpoints}, but this study did not consider any backgrounds.  A second analysis of SUGRA models with large $\tanbeta$ and large \dm\ ($\sim$50~GeV)  showed that high \Et\ hadronic $\tau$'s can be used to reconstruct final states at the LHC \cite{large_dm}.  It has also been shown that a small \dm\ can be measured at the ILC to 10\% \cite{ilc}, but the LHC will be available long before the ILC.  An analysis similar to this one has shown that it is possible to measure a small \dm\ at the LHC to 18\% with a luminosity of 10~fb$^{-1}$, but it was assumed there that an independent gluino mass measurement would exist at the 5\% level \cite{two_tau}.  While this may be a good assumption, at high $\tanbeta$ and small \dm, low energy final state $\tau$'s can complicate the interpretation of the typical \mg\ measurment and may affect the measurement uncertainty as well as the reliability \cite{gluino}.  

\mysect{Roadmap to our paper}

In this letter, we describe new techniques and the prospects of measuring \dm\ and \mg\  simultaneously at the LHC, helping  to establish whether we reside in the co-annihilation region. Following \cite{ilc,two_tau}, we consider an mSUGRA scenario in the co-annihilation region with $\tanbeta$~=~40~GeV, $A_{0}$~=~0, and \mbox{$\mu>0$}, but we allow $m_{0}$ and $m_{1/2}$ to vary such that \dm\ $<$ 20~GeV.  In Sec. 2, we describe the requirements to select inclusive 3$\tau$+jet+$\mett$ events and discuss the differences between this 3$\tau$ analysis and a similar  analysis done with two $\tau$'s.  In particular, we motivate our selection criteria and  describe the mass and counting observables.  In this section, it will become clear that a primary experimental requirement for this analysis to be realized in practice is the efficiency to identify taus with \Et\ as low as 20~GeV.  In Sec. 3, we discuss a method to use both observables to simultaneously determine both \dm\ and \mg.  Conclusions are given in Sec. 4.


\section{Detecting a SUSY Signal in the Co-annihilation Region Using  the 3$\tau$ Final State}


\mysect{SUSY production at LHC}

The primary SUSY production processes at the LHC are $pp$\goes \squark \gluino, \squark \squark, \gluino \gluino.  In each case they decay via \squark \goes $q'$\cone\ or $q$\ntwo\ (or \mbox{\squarkr\goes$q$\none}); \mbox{\gluino\goes$q\bar{q}'$\cone} or $q\bar{q}$\ntwo; and \mbox{\gluino\goes\tbar\tops} or \bbar \sbottom\ and their charge conjugate states, generally producing high \ett\ jets and gaugino pairs.  Since we are interested in the \mbox{\stau\ - \none} mass difference, we must investigate the \mbox{$\stau\goes\tau\none$} decay.  The branching ratio of \mbox{\ntwo\goes$\tau$\stau} is about 97\% for our parameter space and is dominant even for large $m_{1/2}$ in the entire co-annihilation region.  The same is true for the \mbox{\cone\goes$\nu$\stau} decay mode. Therefore, the \mbox{$\stau\goes\tau\none$} decay is present in events with  \ntwo\ntwo, \cone\ntwo, \CONEP\CONEM, and \ntwo\none\ pairs. The decays of \cone\ particles are less desirable than the decays of \ntwo\ particles even though the decay of squarks and gluinos have a larger branching fraction to \cone's than to \ntwo's.  The reason is that \ntwo\ decays produce a correlated pair of one high energy tau from \mbox{\ntwo \goes $\tau$\stau} and one low energy tau from \mbox{\stau \goes $\tau$\none} from which we can extract kinematical information (\mbox{\ntwo\goes$\tau$\stau\goes$\tau\tau$\none}), whereas \cone\ decays only produce a single low energy tau via \mbox{\cone\goes\stau$\nu$\goes$\tau\nu$\none}.  Therefore, we focus on isolating $\tau$ pairs from \ntwo\ decays.  We note that we only use hadronic $\tau$'s since the leptonic decays make reconstruction and $\tau$ identification difficult.

\mysect{final state: 3 tau vs. 2 tau}

The \ntwo\ntwo, \ntwo\cone, and \ntwo\none\ channels each contain a \ntwo\ decay chain and lead to final states with four, three, and two $\tau$'s respectively, each with experimental advantages and disadvantages.  The 4$\tau$ final state is the cleanest in that it has very small background, especially from SM sources; the disadvantage is that it has poor acceptance and efficiency for reconstructing all four $\tau$'s.  The 2$\tau$ final state is the inverse case;  as discussed in \cite{two_tau},  there are significant backgrounds that must be dealt with, but the acceptance is large because a 2$\tau$ signature allows all three channels.  The 3$\tau$ analysis provides a compromise.  We study it here as an alternative to the 2$\tau$ case if the backgrounds for the 2$\tau$ case are underestimated. It will require actual data to determine which analysis performs better or if both are needed.  It may be that the two analyses can be combined to provide additional information, and this is under study.

\mysect{backgrounds}

The 3$\tau$ final state contains both SM and SUSY backgrounds.  The primary SM background is \ttbar\ production.  This produces two $\tau$'s, one from each \mbox{$t$\goes$Wb$\goes$\tau\nu b$}, and a third $\tau$ coming from either a bottom decay (expected to be non-isolated) or from a jet faking a $\tau$.  There are two major sources of SUSY background: real $\tau$'s from decays of \cone 's or \tops 's.  Both of these can produce low or high energy $\tau$'s, but they are uncorrelated and, as we will see, can be separated statistically from \ntwo\ decays by opposite-signed/like-signed subtraction ($OS-LS$).  In addition,  jets faking $\tau$'s leads to additional backgrounds due to both SM and SUSY production, but these can be handled similarly.

\mysect{OS-LS}

Our primary method for separating $\tau$ pairs from \ntwo\ decays from other sources is to use the standard $OS-LS$ technique,  which has been used in other analyses \cite{two_tau,RPV}. To do this in our case,  we order the taus by $\Et$\ ($E_{T}^{\tau_{1}}>E_{T}^{\tau_{2}}>E_{T}^{\tau_{3}}$) and only consider the pairs $\tau_{1}\tau_{3}$ and $\tau_{2}\tau_{3}$.  For illustration, consider a \ntwo\ntwo\ event.  We expect $\tau_{1}$ and $\tau_{2}$ to be from the two \mbox{\ntwo\goes$\tau$\stau} decays because the mass difference between the \ntwo\ and the \stau\ is large ($M_{\ntwo}\approx$267~GeV and $M_{\stau}\approx$154~GeV), and we expect $\tau_{3}$ to be from the  \stau \goes $\tau$\none\ decay, since the mass difference between the \stau\ and the \none\ is small ($M_{\none}\approx$144~GeV).  We would expect a $\tau_{4}$ to also come from the \mbox{\stau \goes $\tau$ \none} decay, but we do not consider it in this analysis. Thus, either $\tau_{1}\tau_{3}$ or $\tau_{2}\tau_{3}$ is from a \ntwo, so the taus of one pair are correlated and are opposite-signed ($OS$).  In the other pair the $\tau$'s are uncorrelated and can be $OS$ or $LS$  with equal probability.  Therefore, we subtract the number of observed $LS$ pairs from the number of $OS$ pairs and only use pairs in the kinematically allowed region. As a second example, if our event contains a \ntwo\ and a fake $\tau$ from a jet from squark decay, we will have two high energy $\tau$'s and one low energy $\tau$.  Again, either the $\tau_{1}\tau_{3}$ or $\tau_{2}\tau_{3}$ pair is from the \ntwo, producing a pair that is correlated and $OS$, while the other pair is uncorrelated and has equal probability of being $OS$ or $LS$.  This same method works for any background where a \ntwo\ is produced along with a $\tau$ from another source, for example, from \cone\ decay.

\mysect{Event Generation}

To generate our data sets, we simulate our model with all SUSY production using ISAJET \cite{isajet}. We run the generated particles through a detector simulator, PGS \cite{pgs}, using the CDF parameter file for jet finding, and directly use the generated particles for the non-jet objects.  Finally, a separate Monte Carlo routine  simulates the effects of efficiency for $\tau$'s  and fake rate for jets.  Based on CDF results and preliminary LHC experimentally studies, we take the $\tau$ efficiency to be $50\%$ for $\tau$'s with $\Et >$~20~GeV and the $\tau$ fake rate for a jet to be \mbox{$1\% \pm 0.2\%$} \cite{jets}.

\mysect{tau Et cuts}

Based on the simulations, the full set of selection criteria are listed in Table~\ref{criteria}.  The first two $\tau$'s have high enough average energies that it is reasonable to cut at \Et$>$40~GeV and $|\eta|<2.5$.  At this energy, we can expect efficient identification.  For small \dm\ the energy of the third $\tau$, from the \mbox{\stau\goes$\tau$\none} decay, can be very low, often with $E_{T}$~$\lesssim$~15~GeV.  As in \cite{two_tau}, we loosen the \Et\ requirement to 20~GeV as a balance between keeping the maximum number of events while still being in a region of reasonable reproduction of the $\tau$ identification capabilities at the LHC detectors. 

\mysect{jet met cuts}

To reduce both SUSY and SM backgrounds, we require at least one jet and \mett. Since we expect all events to end in two \none's, we require the \mett\ to be large. In addition, as previously mentioned,  squark decays result in high energy jets.  We impose the cuts \mbox{\mett$>$100 GeV}, \mbox{$E_{T}^{\mathrm{jet}~1}$ $>$ 100 GeV} ($|\eta|<$2.5), and \mbox{\mett$ + E_{T}^{\mathrm{jet}~1}$ $>$ 400 GeV}.  We require cuts on the jet and \mett\ separately to insure that SM events with unusually high jet $E_{T}$ or \mett\ have a low likelihood of passing cuts.  These cuts have the further advantage that events passing them should readily pass reasonable \mbox{jet+\mett} triggers envisioned for the LHC.

\mysect{invariant mass cuts}

The invariant mass of the $\tau$ pair from the \ntwo\ decay forms a mass peak and provides excellent rejection against both backgrounds \cite{endpoints, two_tau}. This can be seen by considering the $\tau$ pair in the chain \mbox{\ntwo\goes$\tau$\stau\goes$\tau\tau$\none} in the rest frame of the \ntwo.  In this case, the invariant mass is given by

\begin{equation}
M_{\tau\tau} = M_{\ntwo}\sqrt{\frac{1-\mathrm{cos}\theta}{2}}\sqrt{1-\frac{M^{2}_{\stau}}{M^{2}_{\ntwo}}}\sqrt{1-\frac{M^{2}_{\none}}{M^{2}_{\stau}}}
\end{equation}

\noindent where $\theta$ is the opening angle between the $\tau$'s \cite{endpoints}; the kinematic endpoint of this distribution corresponds to  the $\theta$ = $\pi$ case.  Therefore, the peak and endpoint depends mostly on $M_{\ntwo}$, $M_{\stau}$, and $M_{\none}$, of which $M_{\ntwo}$ and $M_{\none}$ depend heavily on \mg\ from the mSUGRA relations.  Fig. \ref{MttDist} shows the \mtt\ distribution for SUSY and \ttbar\ production.  We select $\tau$ pairs with $M_{\tau\tau}<$100~GeV because this is just beyond the endpoint for all \dm\ and \mg\ values in our parameter range.   This makes our results less sensitive to the fake rate uncertainty because jets from squarks that fake $\tau$'s tend to have large \Et, which can produce large $M_{\tau\tau}$. In Fig. \ref{MttDist}, we also see that the number of events in the peak of the $OS$ distribution in excess above the $LS$ distribution increases  with increasing \dm.  In addition, the peak position also increases as a function of \dm. Both these features will be used in the next section.  We note that the \ttbar\ distribution also has a peak, but the event rate is several orders of magnitude less than the SUSY production even without standard isolation cuts, and it is ignored throughout the rest of this letter. We also note, that as in Ref.~\cite{two_tau}, lowering the \Et\ requirement of the $\tau_3$ to 20~GeV is needed for the peak to be visible.

\ThreePlot{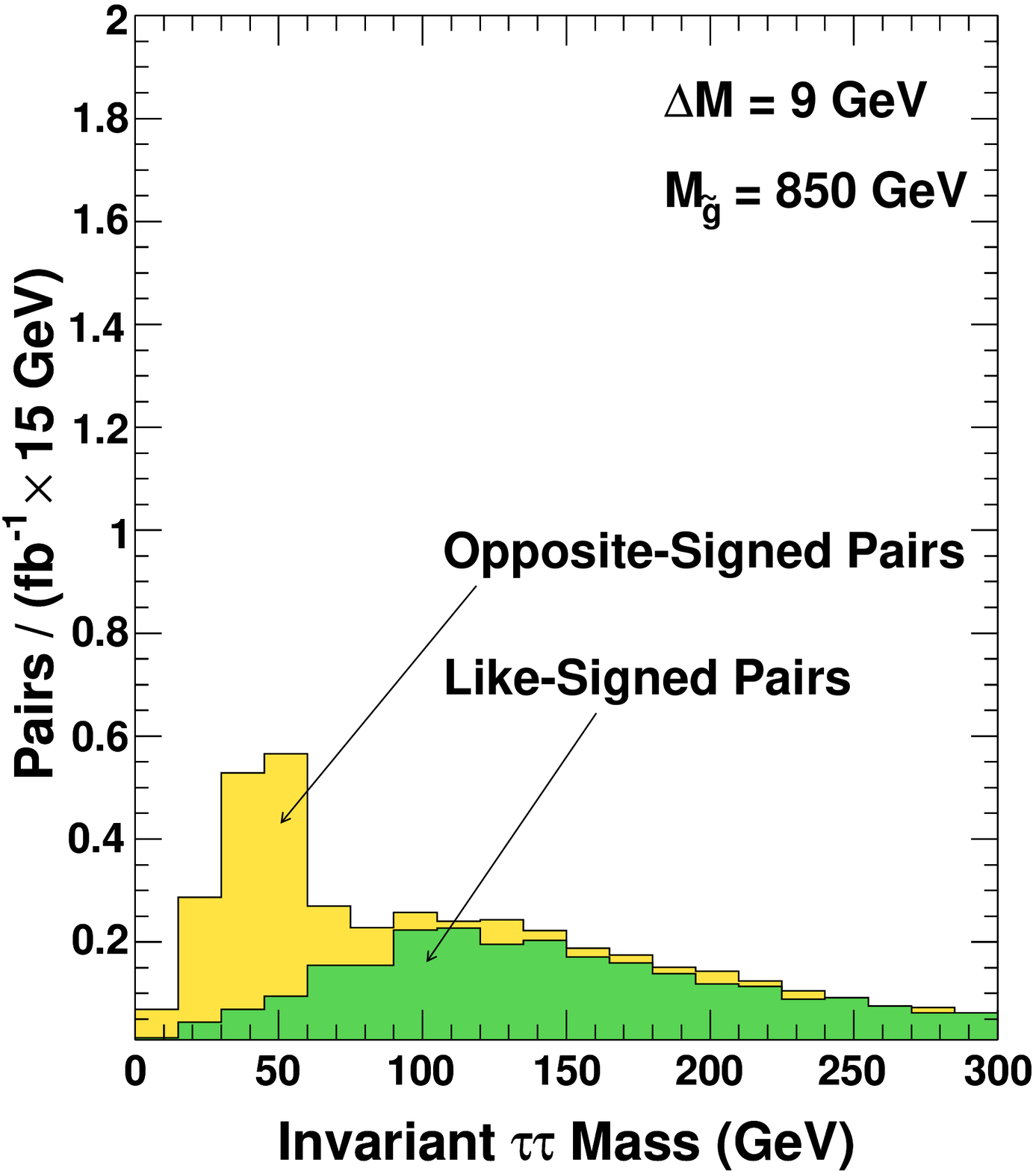}{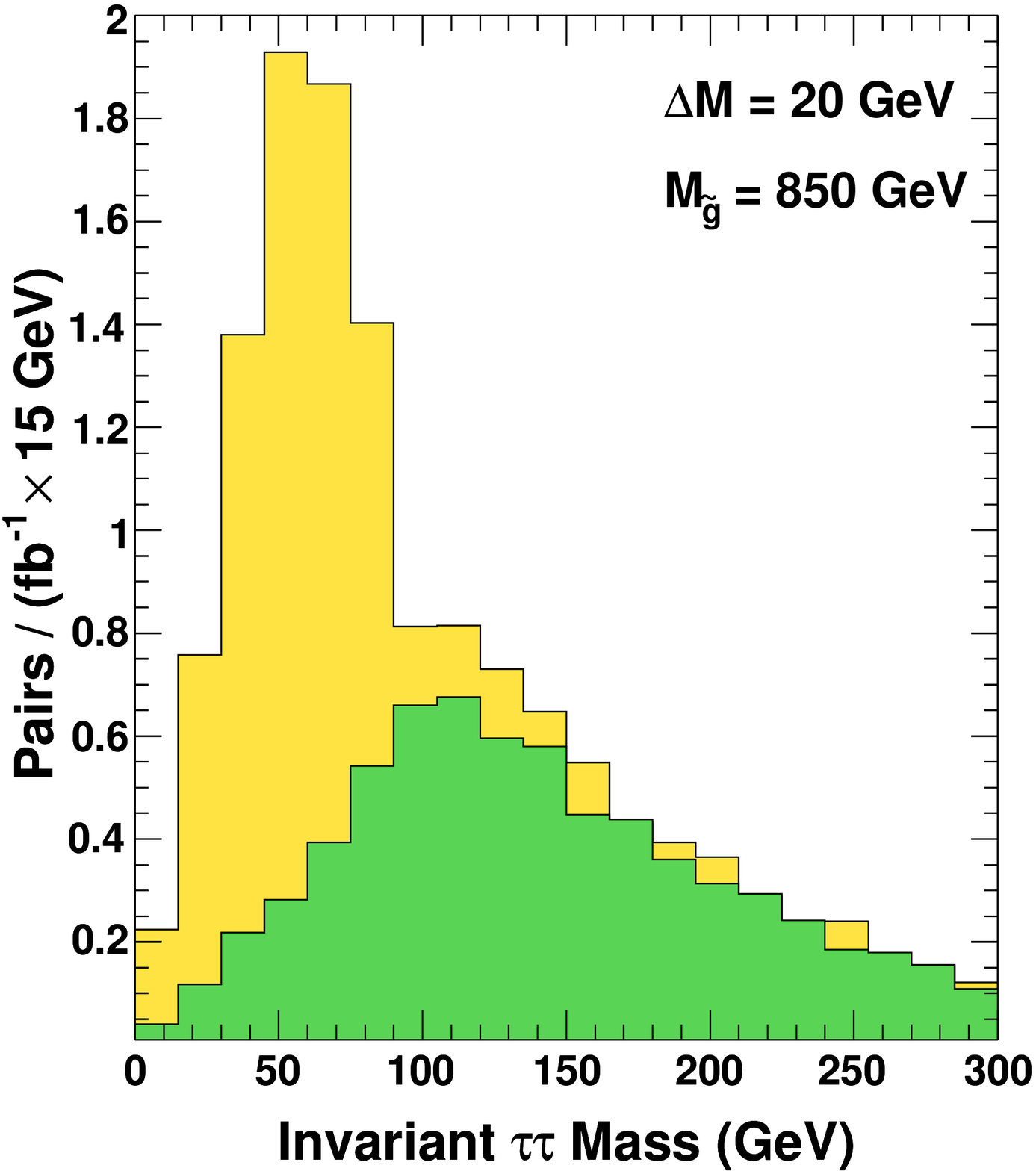}{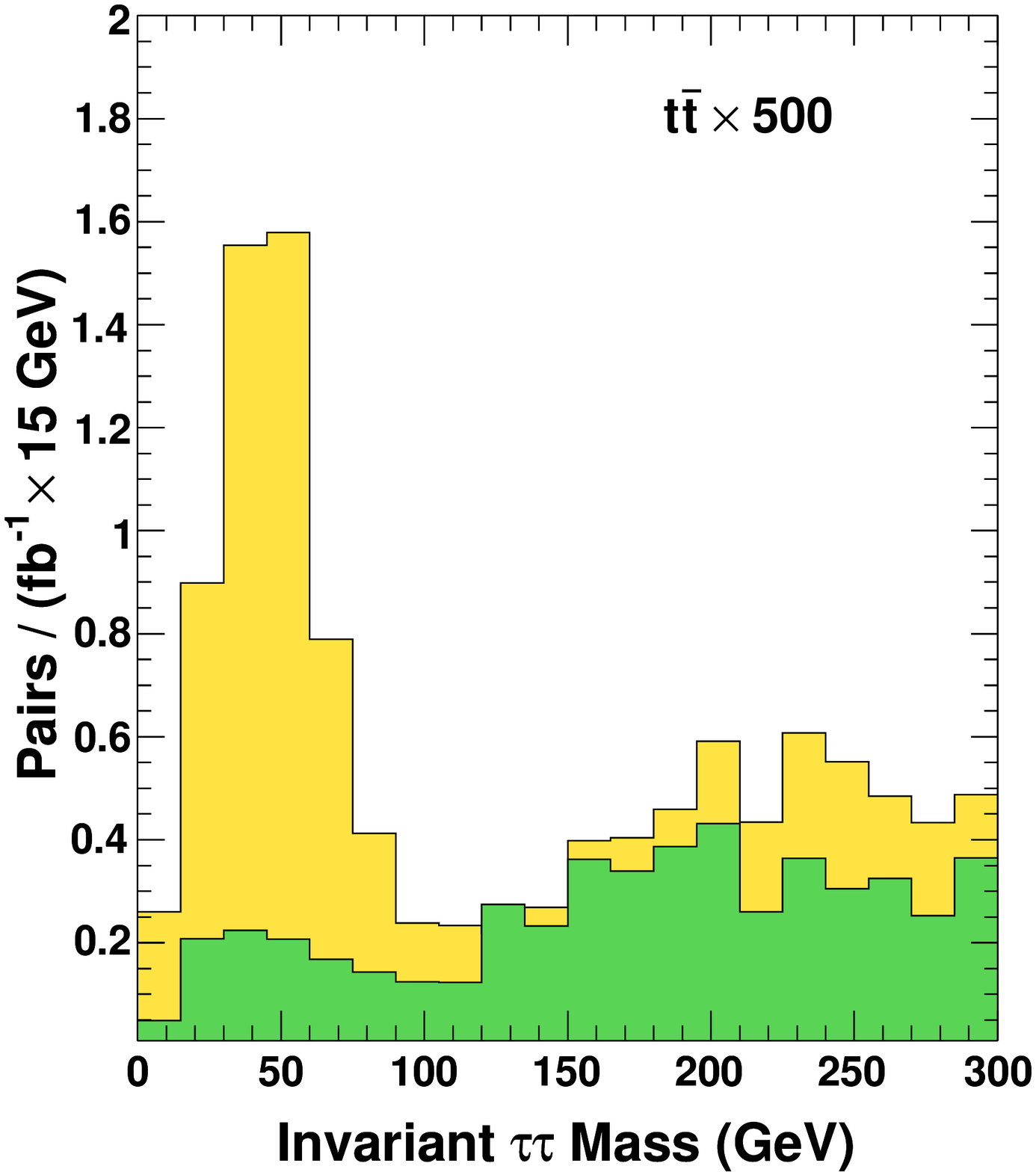}{The $M_{\tau \tau}$ distribution of $OS$ and $LS$ $\tau$ pairs for SUSY events with \dm\ = 9~GeV (top left) and \dm\ = 20~GeV (top right),  and $\ttbar$ background (bottom).  The $LS$ distribution has a characteristic shape of uncorrelated particles.  Note that subtracting the  $LS$ distribution from the $OS$ distribution leaves a mass peak due to both $\tau$'s originating from a single \ntwo.  For small \dm, this peak contains few counts and is centered at low mass, and for large \dm, this peak contains many counts centered at a higher mass.  There is always a kinematical cutoff below $\sim$100~GeV.  We note that \ttbar\ production also produces a peak, but the event rate is several orders of magnitude smaller than for SUSY production.  Therefore, we assume that the SM background is negligible.}{MttDist}

\begin{table}
\caption{Final selection criteria.}
\label{criteria}
\begin{center}
\begin{tabular}{l}
\hline
3 identified $\tau$ candidates with $|\eta |$ $<$ 2.5 and $E_T$ $>$ 40, 40 and 20 GeV respectively\\
1 jet with $E_T$ $>$ 100 GeV and $|\eta|$ $<$ 2.5\\
\mett $>$ 100 GeV\\
\mett + $E_T^{\mathrm{jet}~1}$ $>$ 400 GeV\\ 
$M_{\tau\tau}$ $<$ 100 GeV where only $\tau_1\tau_3$ and $\tau_2\tau_3$ invariant mass combinations are considered\\
\hline
\end{tabular}
\end{center}
\end{table}

\section{Simultaneous Measurement of \dm\ and \mg}

\mysect{Intro to measuring}

\indent In this section, we define our observables, the number of counts (\nosls) and the ditau mass peak position (\mtt) and describe their values as a function of both \dm\ and \mg.  We then show how these two variables can be used to simultaneously measure both \dm\ and \mg, and compare our \dm\ result to previous analyses that assume an independent \mg\ measurement.  


\mysect{Counting overview: \nosls\ vs. \dm}

The variable \nosls\ is the number of $LS$ $\tau$ pairs subtracted from the number of $OS$ $\tau$ pairs passing all the selection requirements in Table \ref{criteria}.  Because we expect the $\tau_{3}$ to come from the \mbox{$\stau\goes\tau\none$} decay, the average \Et\ of the $\tau_{3}$, its probability of having $E_{T}^{\tau_{3}}$~$>$~20~GeV, and, therefore, \nosls\  grow with \dm.  Thus, for a known luminosity, a measurement of \nosls\ allows for a determination of \dm.  An increase in \mg\ affects \nosls\ by decreasing the production rate of gluinos, which decreases the number of \ntwo\ decay chains produced.  However, an increase in \mg\ increases the boost of the \ntwo, which increases the number of accepted \ntwo\ decay chains.  The overall effect is to decrease \nosls\ with increasing \mg. Though mass changes in squarks and the neutralinos also modify both production and boost,  these mass effects change with \dm\ and \mg\ in mSUGRA.  Therefore, \mg\ provides a scale for the model. Fig. \ref{Counts_v_Mglu} shows \nosls\  as a function of \dm\ and \mg.  We see that \nosls\ is flat below \dm~$\sim$5~GeV and nearly linear above it.  At low \dm, the number of $\tau$ pairs from single \ntwo\ decays goes to zero as none of the $\tau$'s from \stau\ decay pass the 20~GeV threshold; however, \nosls\ does not go to zero because of the small SUSY background from stop quark pair production and decay via \mbox{$\tops \goes t\widetilde{\chi}^0_i \goes (Wb)\widetilde{\chi}^{0}_{i} \goes (\tau\nu)b\widetilde{\chi}^{0}_{i}$}. This background is independent of \dm, so the small number of events in the very low \dm\ region implies that it is negligible.

\TwoPlot{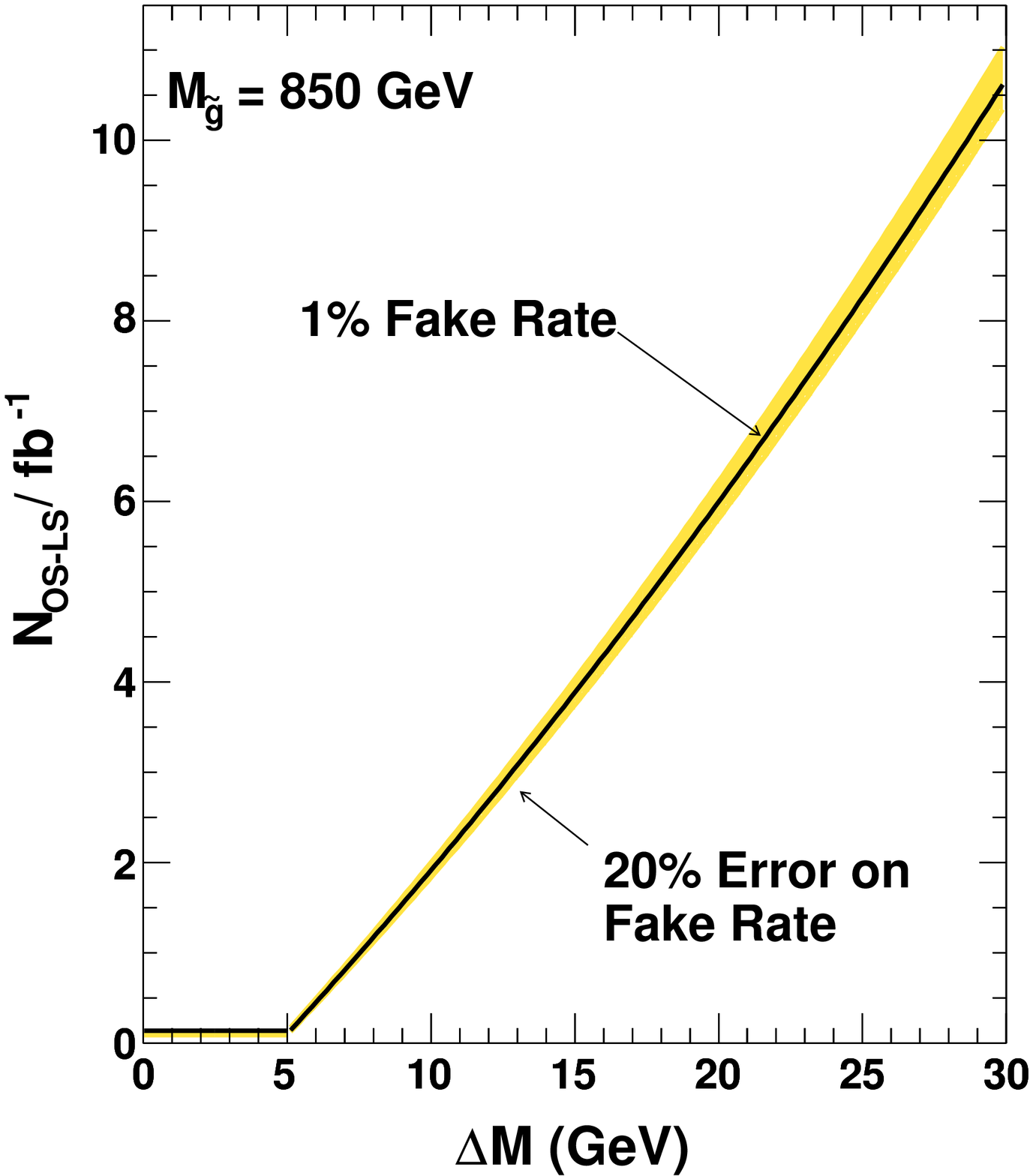}{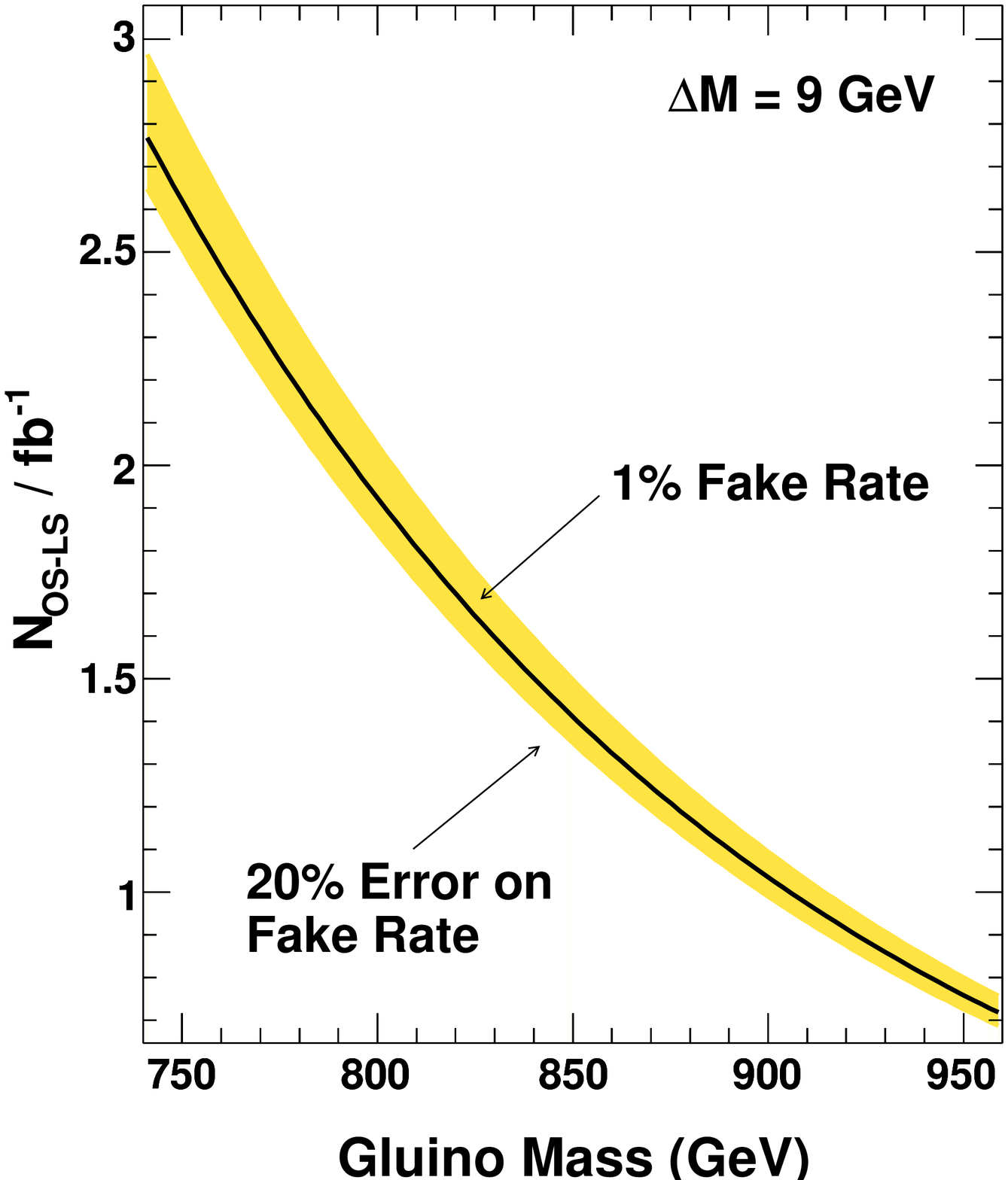}{The left plot shows \nosls\ as a function of \dm\ with a 1\% fake rate with the shaded band representing the variation due to the 20\% systematic uncertainty on the $\tau$ fake rate. Below $\sim$5 GeV, the third $\tau$ from \mbox{\stau\goes$\tau$\none} is so soft that there is no signal; therefore, counting is dominated by SUSY backgrounds, and the number of counts is flat.  Above $\sim$5 GeV, the number of counts is nearly linear as a function of \dm\ as more and more $\tau$'s from \mbox{\stau\goes$\tau$\none} pass the 20~GeV threshold.  The right plot shows the relationship between \nosls\ and \mg\ for \dm\ =~9~GeV;  \nosls\ decreases strongly with increasing \mg.}{Counts_v_Mglu}

\mysect{Uncertainty}

We consider two contributions to the uncertainty in the \nosls\ measurement: the statistical uncertainty and the uncertainty in the fake rate.  We note that statistical uncertainty is not the usual $\sqrt{N}$ but, from binomial statistics, $2\sqrt{\frac{N_{OS}N_{LS}}{N_{OS}+N_{LS}}}$.  The 20\% uncertainty in the $\tau$ fake rate also produces a systematic uncertainty on the true value of \nosls.  We find the uncertainty due to the fake rate to be small compared to the statistical uncertainty even though  $\sim$20\% of the \nosls\ event rate is due to events with at least one jet faking a $\tau$.  While this may seem puzzling, because of the $OS-LS$ procedure, the additional counts due to the fake rate predominantly come from situations where a $\tau$ pair is produced by a \ntwo\ decay, and the first or second $\tau$ is due to a jet faking a $\tau$.  In this situation, the fake rate is bringing $\tau$ pairs from \ntwo\ decay that would have been included in the 2$\tau$ analysis into our sample; therefore, these additional counts do not reduce our sensitivity.


\mysect{MM overview: \mtt\ vs. \dm}

We define \mtt\ as the position of the peak of the ditau invariant mass distribution after the $LS$ distribution is subtracted from the $OS$ distribution. It directly depends on \dm\ (as shown in Fig. \ref{MttDist}) and indirectly depends on \mg.  The dependences of \mtt\ enter through Eq. 1. Because of the dependence of the formula  on $M_{\stau}$ and $M_{\none}$, \mtt\ rises as a function of \dm\ (shown in Fig. \ref{Peak_v_Mglu}, top left); however, the dependence of \mtt\ on \mg\ comes indirectly because of the mSUGRA relations: \mg\ $\simeq$ 2.8$m_{1/2}$, $M_{\ntwo}$ $\simeq$ 0.8$m_{1/2}$, and  $M_{\none}$ $\simeq$ 0.4$m_{1/2}$.  Therefore, as \mg\ changes, so do $M_{\ntwo}$ and $M_{\none}$, which leads to \mtt\ rising as a function of \mg.  This is shown in Fig.~\ref{Peak_v_Mglu}, top right.

\FourPlot{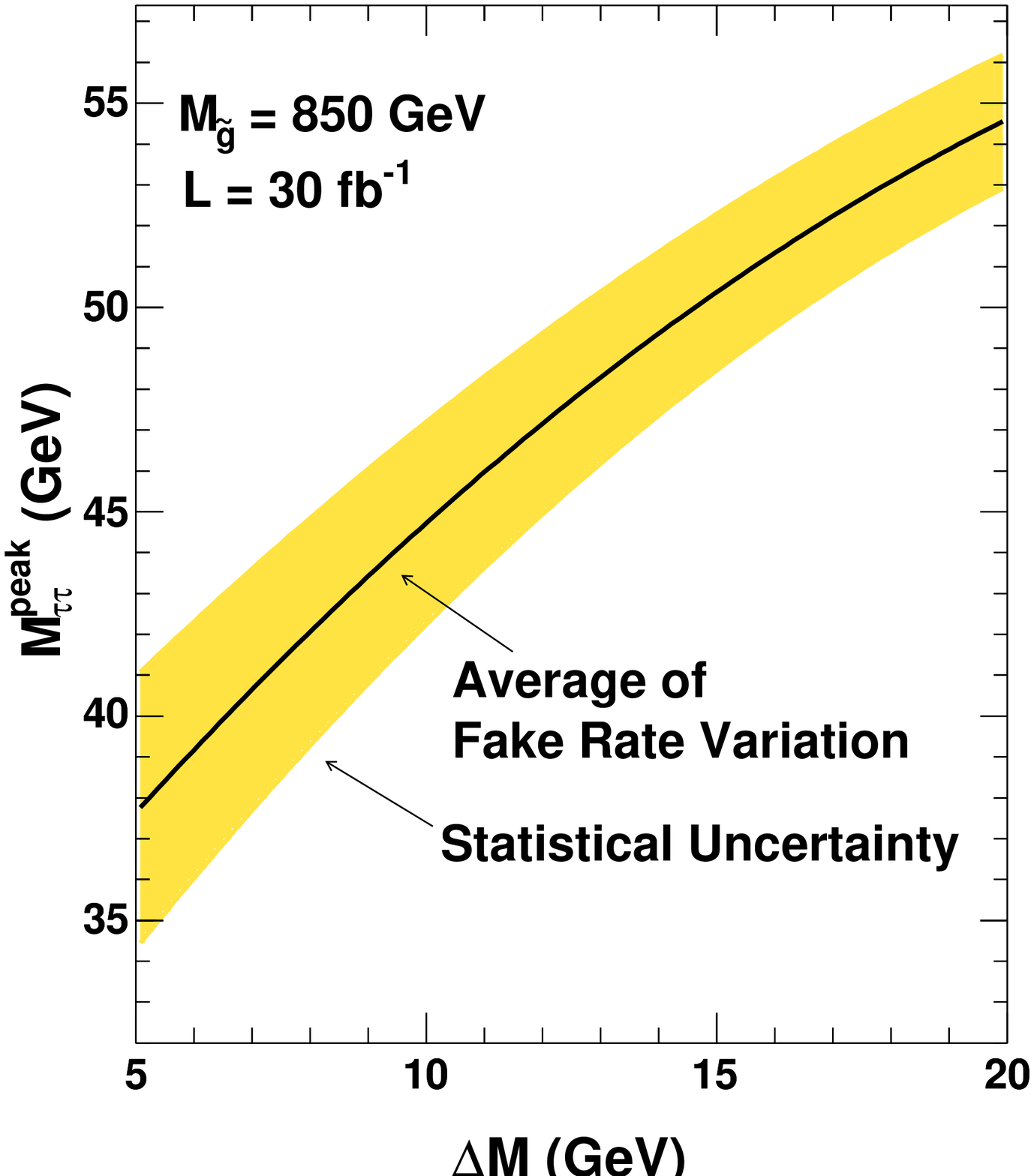}{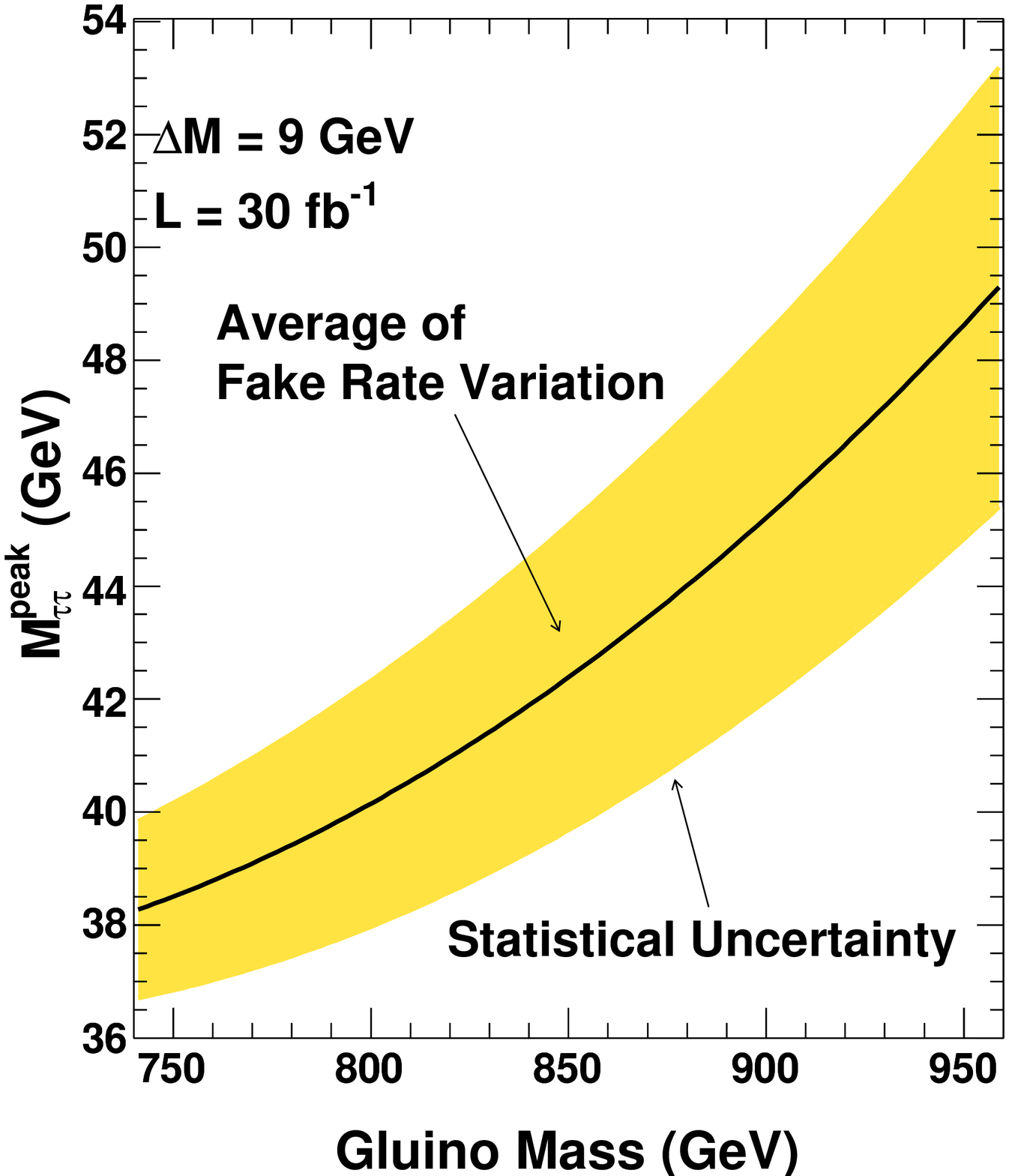}{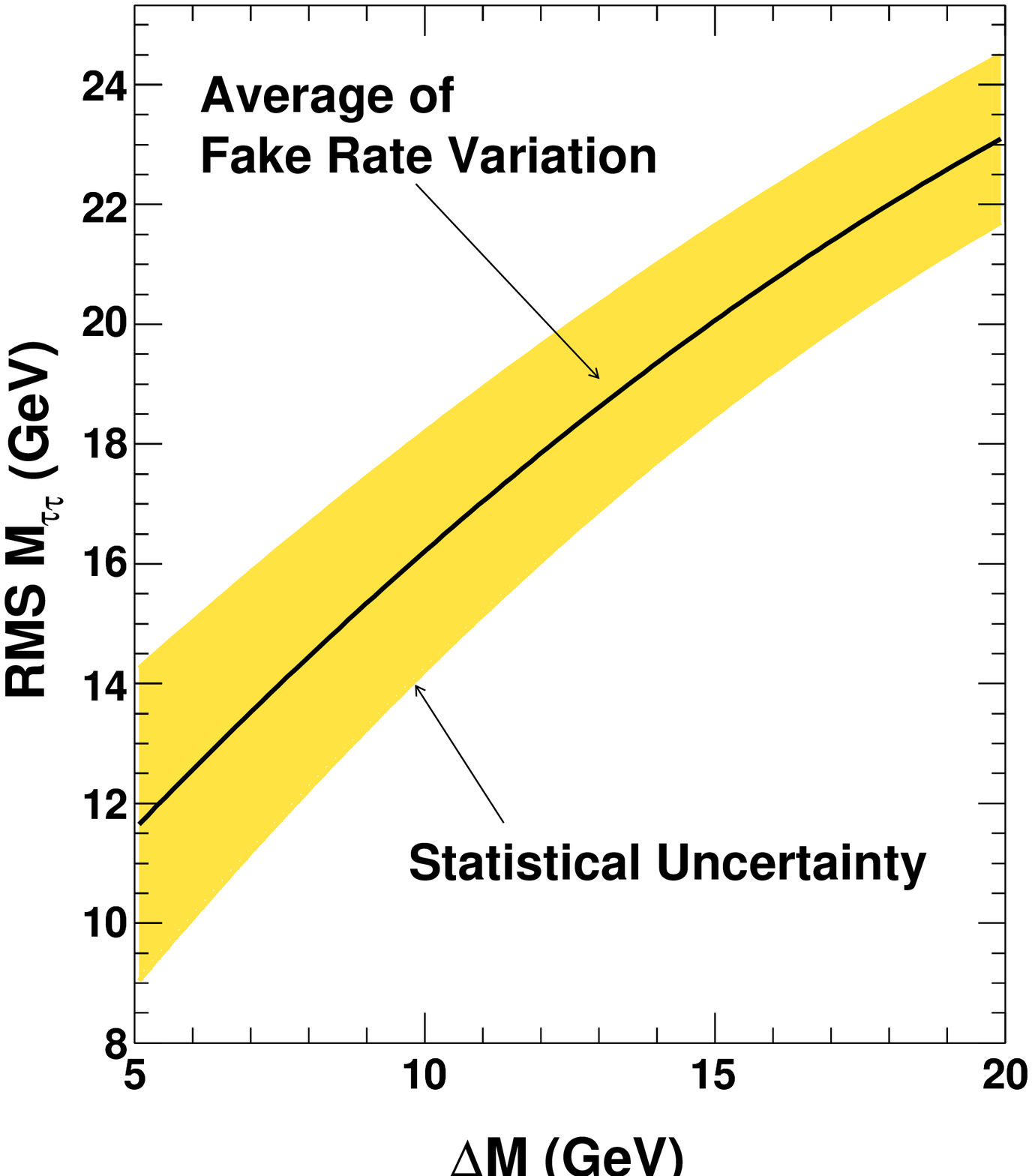}{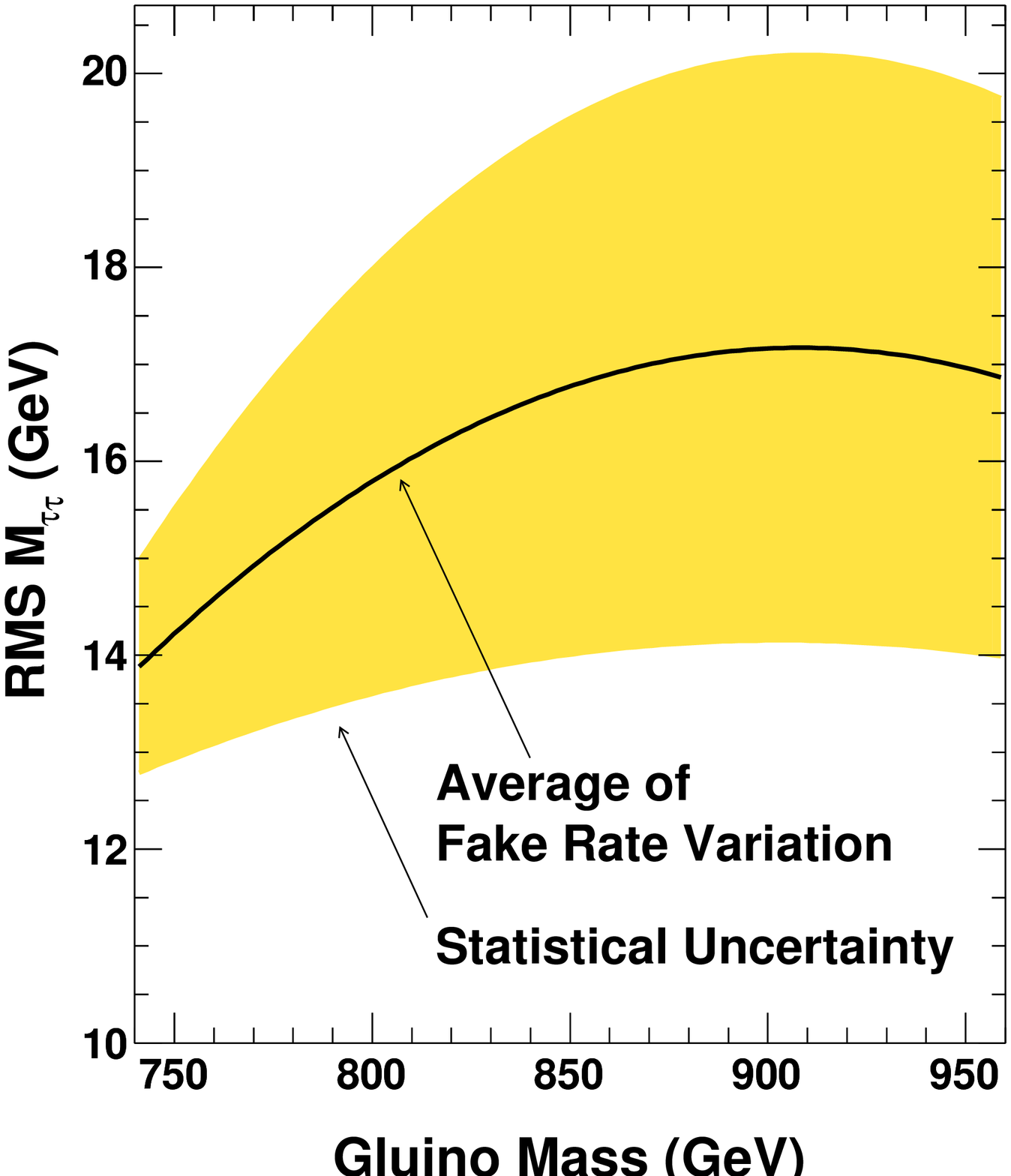}{The upper plots show \mtt\ as a function of \dm\ (left) and \mg\ (right).  The lower plots show the RMS of the $M_{\tau\tau}$ distribution as a function \dm\ (left) and \mg\ (right). The statistical uncertainty is significantly larger than the uncertainty due to fake rate variation and is taken here normalized to expectations for 30~fb$^{-1}$.  Note that the mass method is not possible for \dm\ values less than 5 GeV due to the inability to form a mass peak even at high luminosities.  The \mtt\ increases strongly with both increasing \dm\ and \mg.}{Peak_v_Mglu}

\mysect{MM statistical uncertainty}

Unlike \nosls\ the \mtt\ measurement is dominated by the uncertainty on the number of events in the sample until very high luminosities. We take the uncertainty on \mtt\ to be given by the standard formula, $\sigma_{\mtt} = \mathrm{RMS}/\sqrt{N}$, where the RMS of the mass distribution is shown in Figs. \ref{Peak_v_Mglu} bottom left and right, and $\sqrt{N}$ is the number of counts in the peak (effectively \nosls), which is a function of the luminosity. For $\mathcal{L}$ = 30~fb$^{-1}$ and \mg\ = 850~GeV, we show the statistical uncertainty on \mtt\ as a band in Fig.~\ref{Peak_v_Mglu}. The uncertainty due to the fake rate is unmeasurably small compared to the statistical uncertainty, even for large fake rate values, up to 5\%.  This helps confirm our previous assertion that only $\tau$ pairs from a \ntwo\ survive the $OS-LS$ procedure.  Pairs not from \ntwo\ decays would have an \mtt\ that is characteristically shifted. Since this does not happen, even with high fake rates, we conclude that the additional events are from \ntwo\none\ or \ntwo\cone\ events, which contain $\tau$ pairs from a \ntwo\ and an additional $\tau$ from a fake or a \cone.  This is confirmed by the Monte Carlo.



\mysect{Method}

Since \nosls\ and \mtt\ both depend on \dm\ and \mg, we can invert these dependences to measure \dm\ and \mg\ simultaneously.  To do this, we parameterize \nosls\ and \mtt\ as functions of \dm\ and \mg.  The contours of constant \nosls\ and constant \mtt\ are shown in Fig.~\ref{Contour} for \dm\ = 9~GeV and \mg\ = 850~GeV.  The intersection of these central contours provides the measurement of \dm\ and \mg, and the auxiliary lines, from  expected 1$\sigma$ uncertainties on \nosls\ and \mtt,  respectively, determine the 1$\sigma$ region for \dm\ and \mg. Note that we require the measurement of \nosls\ to be statistically significant in order to make a measurement.  We find that for values of \dm~$\gtrsim$~8~GeV, we require less than 10~fb$^{-1}$ for 3$\sigma$ significance.  For \dm\ below 6~GeV, no amount of luminosity suffices to reach the 3$\sigma$ level since the third $\tau$ is not energetic enough to be observed, and no \mtt\ can be constructed.

\Plot{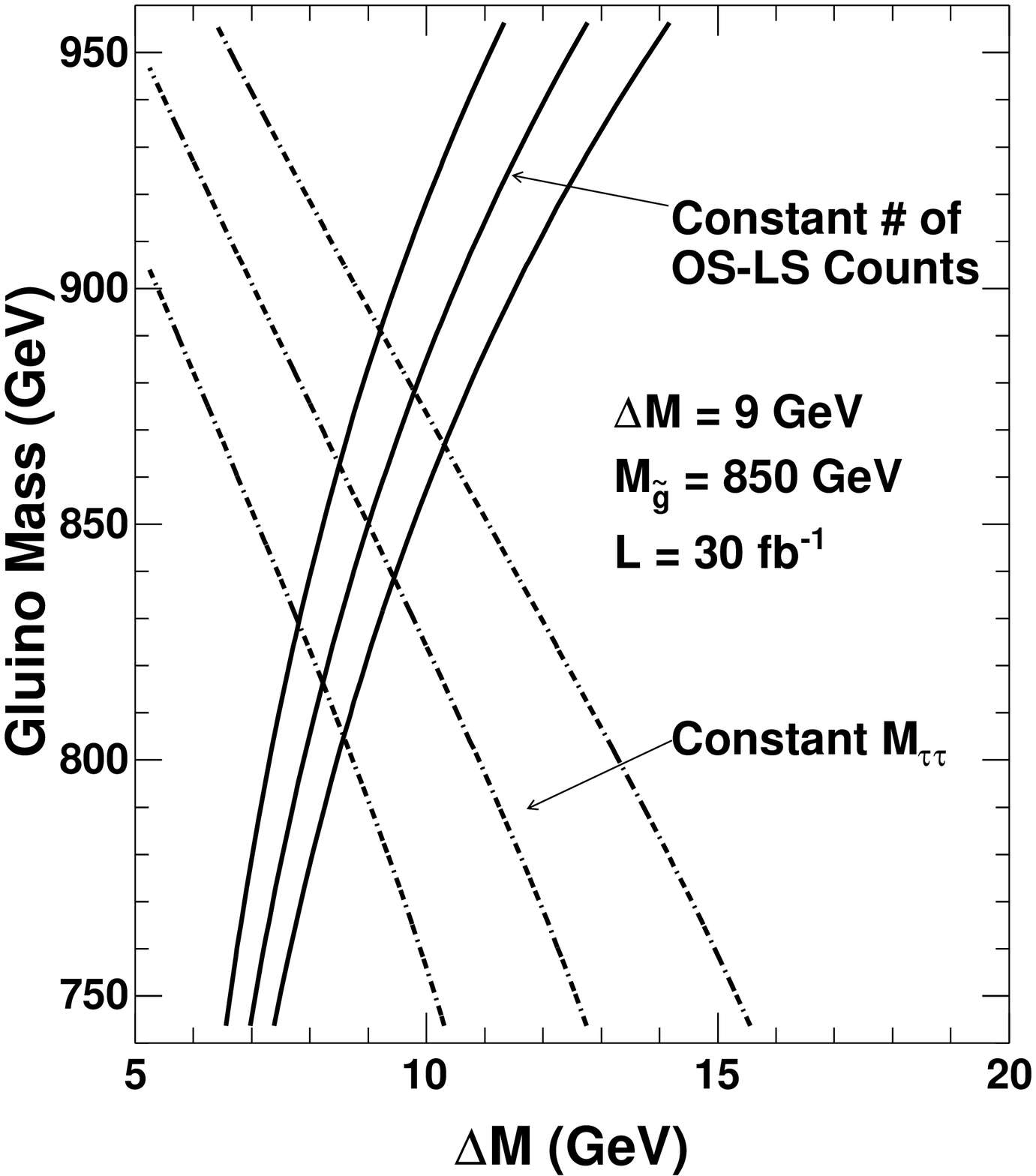}{The contours of constant \nosls\  and \mtt\ for \dm\ = 9~GeV, \mg\ = 850~GeV, and $\mathcal{L}$ = 30~fb$^{-1}$. The middle lines are the central values while the outer lines show the 1$\sigma$ uncertainty on the measurements.  The region defined by the outer four lines indicates the 1$\sigma$ region for the \dm\ and \mg\ measurements.}{Contour}

\mysect{result}

Fig.~\ref{sig_gluino} shows the expected uncertainty on \dm\ as a function of \dm\ and the expected percent uncertainty on \dm\ and \mg\ as functions of luminosity for \dm\ = 9~GeV and \mg\ = 850~GeV. We find that for $\mathcal{L}$ = 30~fb$^{-1}$, we can measure \dm\ to $\sim$15\% and \mg\ to $\sim$6\%.  It is important to note again, however, that our determination of \mg\ is not a direct measurement, but a determination of a parameter in our model, in some sense, the SUSY mass scale of the model.  It will need to be compared to a direct \mg\ measurement, assuming one is available.  If the two results were consistent, it  would be a consistency check of the gaugino universality of the mSUGRA model and that we are indeed in the co-annihilation region.

\TwoPlot{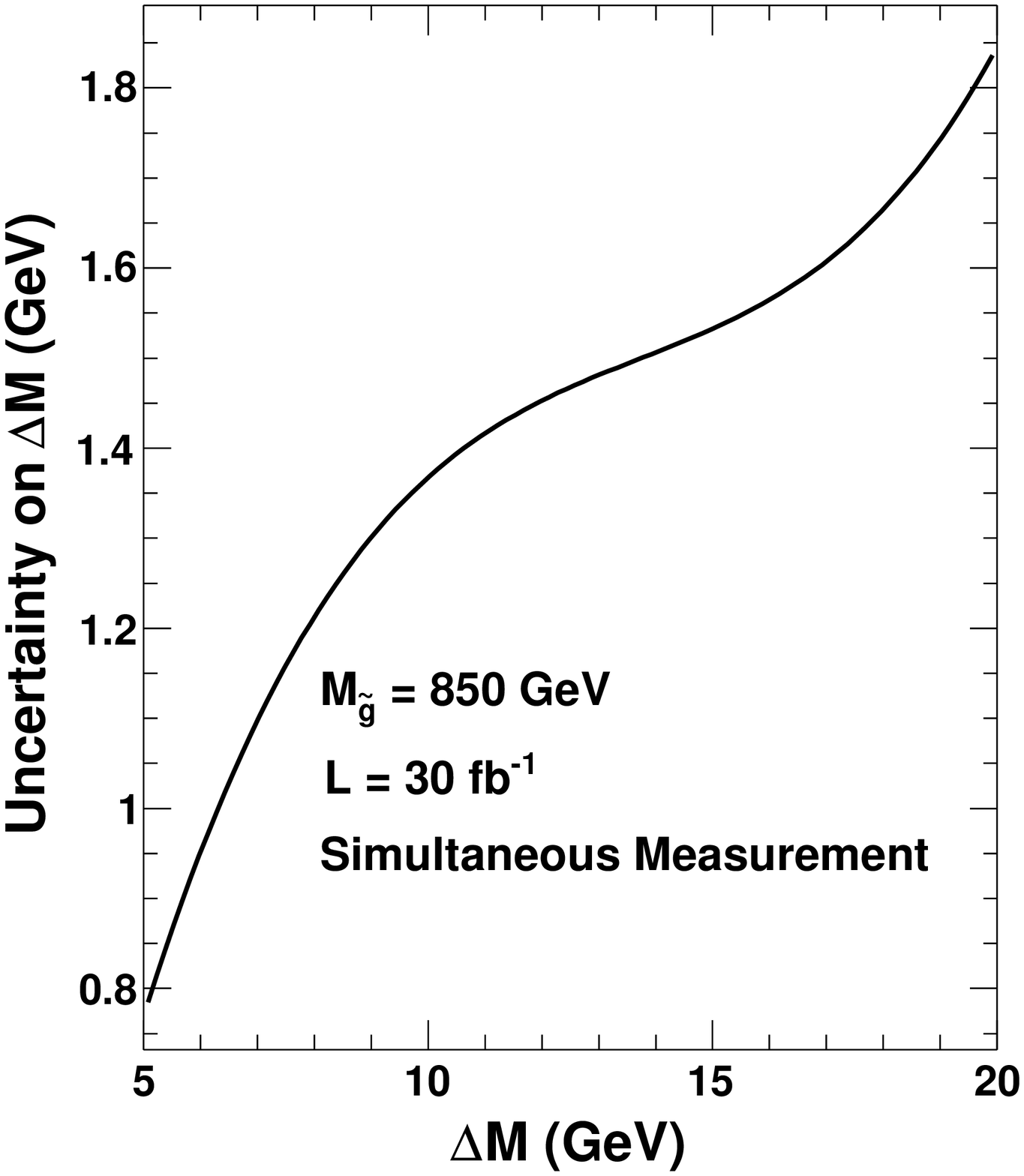}{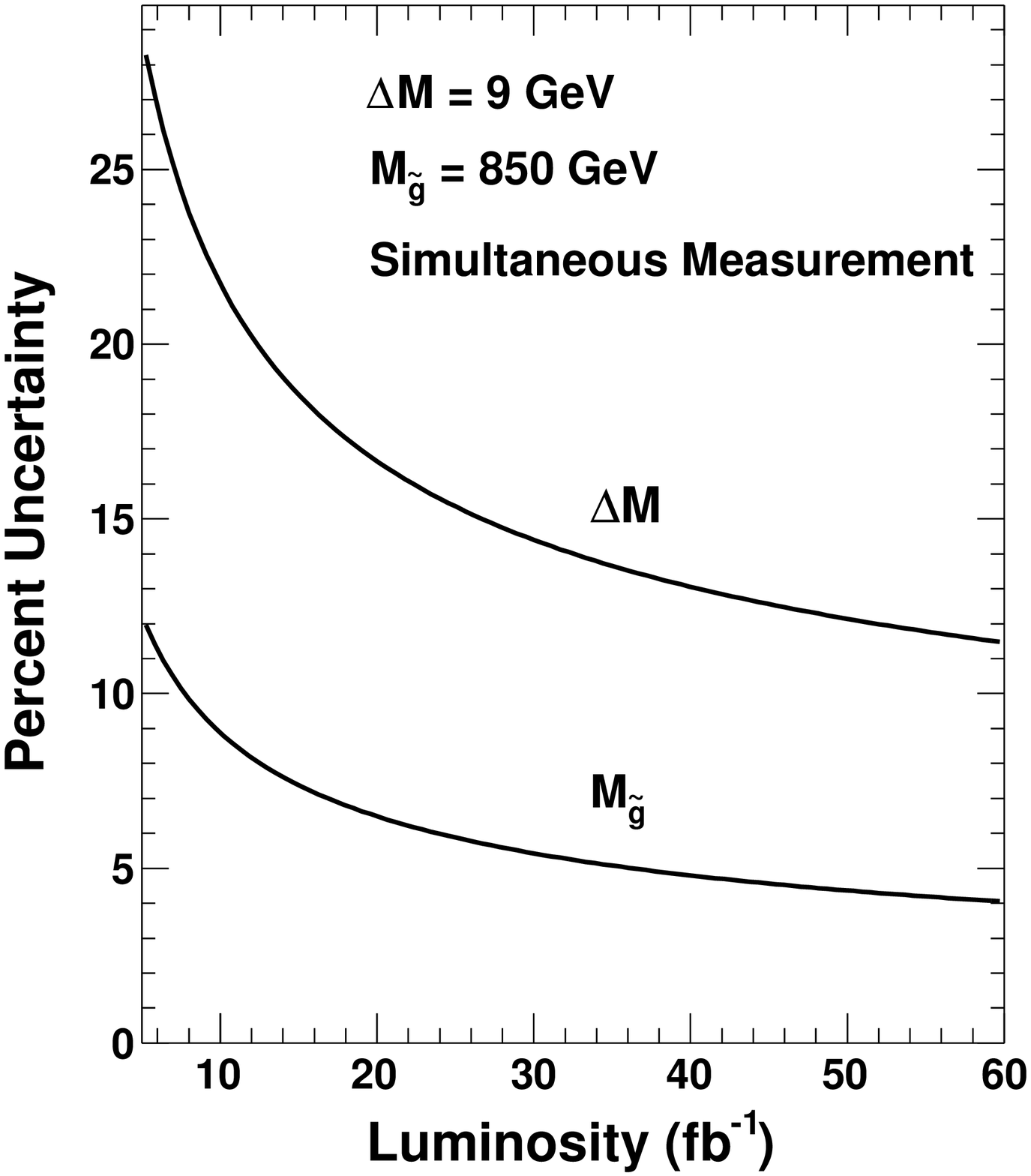}{The expected uncertainty on \dm\ and \mg\ for the simultaneous measurement method. The plot on the left shows the uncertainty on \dm\ as a function of \dm.   The plot on the right shows the percent uncertainty in \dm\ and \mg\ as functions of luminosity for \dm\ = 9~GeV and \mg\ = 850~GeV. At high luminosities, the uncertainties on both \dm\ and \mg\ continue to improve and are statistics limited.}{sig_gluino}



\mysect{measuring \dm}

If we assume an independently measured \mg, as in previous analyses \cite{ilc,two_tau}, our two observables can be considered to be two independent measurements of \dm\ that can be compared and combined for further testing of the model.  We assume here that \mg=850~GeV and has been measured to 5\%, and we incorporate this uncertainty into our \dm\ measurement as a systematic uncertainty.  To do this, we determine the variation in our observables from the expected values at a fixed \dm\ but with \mg\ varying by 5\%.  We then propagate these variations using parameterizations of our observables with \mg\ fixed as in \cite{two_tau}.  The results are shown in the top plots of Fig.~\ref{Sigma_All}.  We see that for both methods, the systematic uncertainty clearly dominates.  Combining the two results, using $\sigma_{combined} = \frac{1}{\sqrt{\sigma_{counting}^{-2}+\sigma_{mass}^{-2}}}$, we find that with $\mathcal{L}$ = 30~fb$^{-1}$ \dm\ can be measured to 12\% at our 9~GeV mass point.  The mass method is always slightly worse than the counting method but does help improve the combined uncertainty.  The full results are shown in the bottom left and bottom right of Fig. \ref{Sigma_All}.

\FourPlot{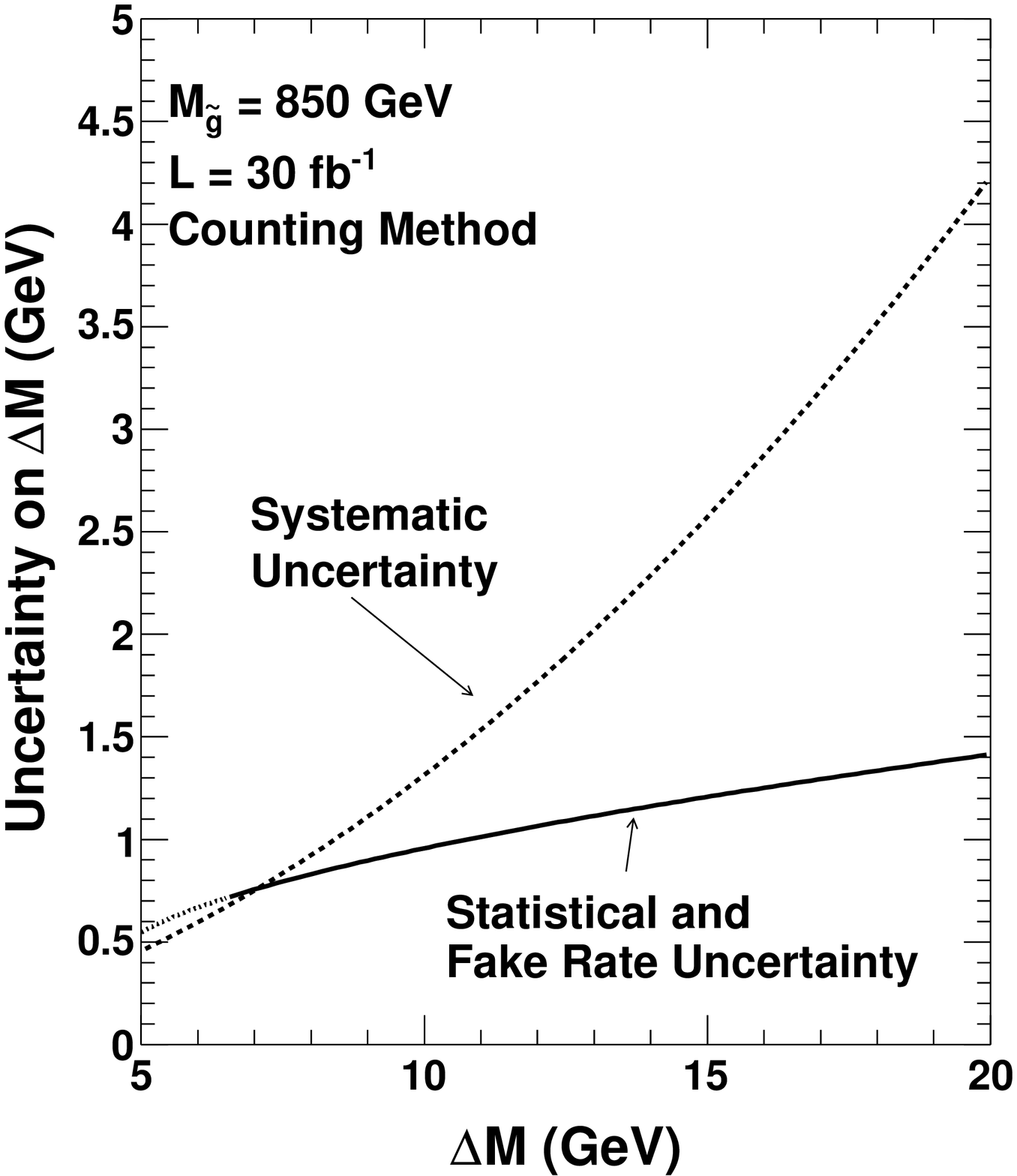}{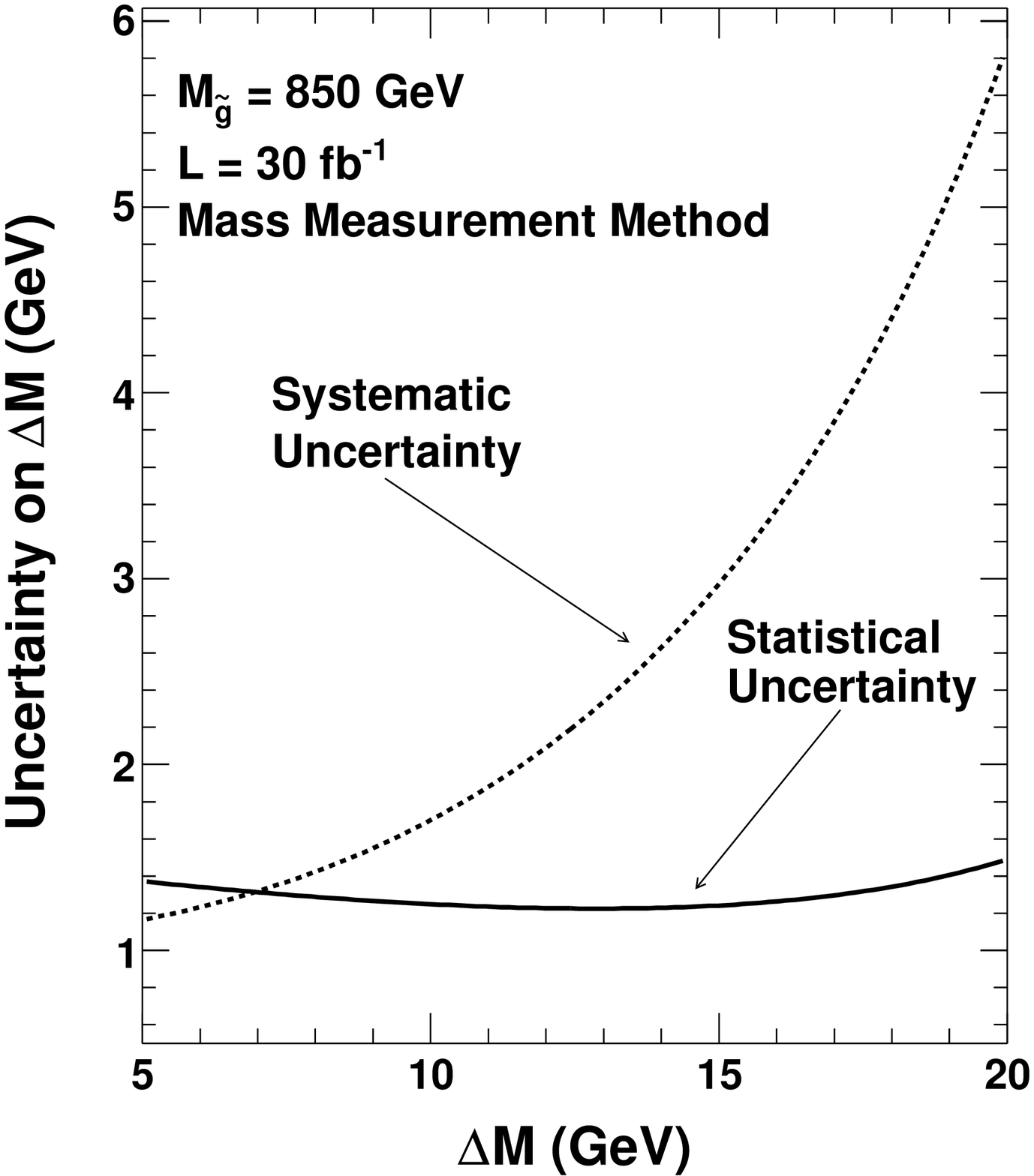}{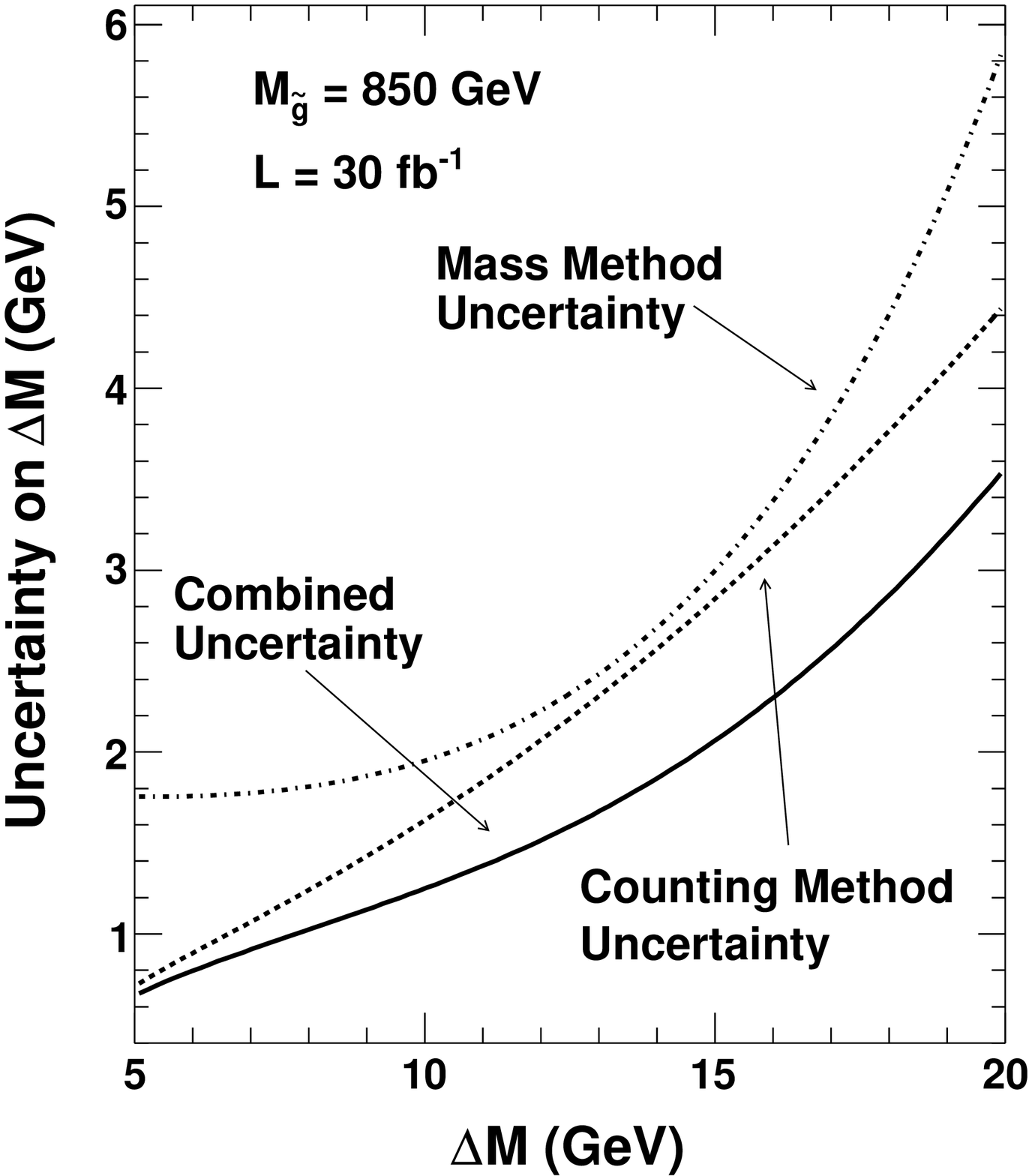}{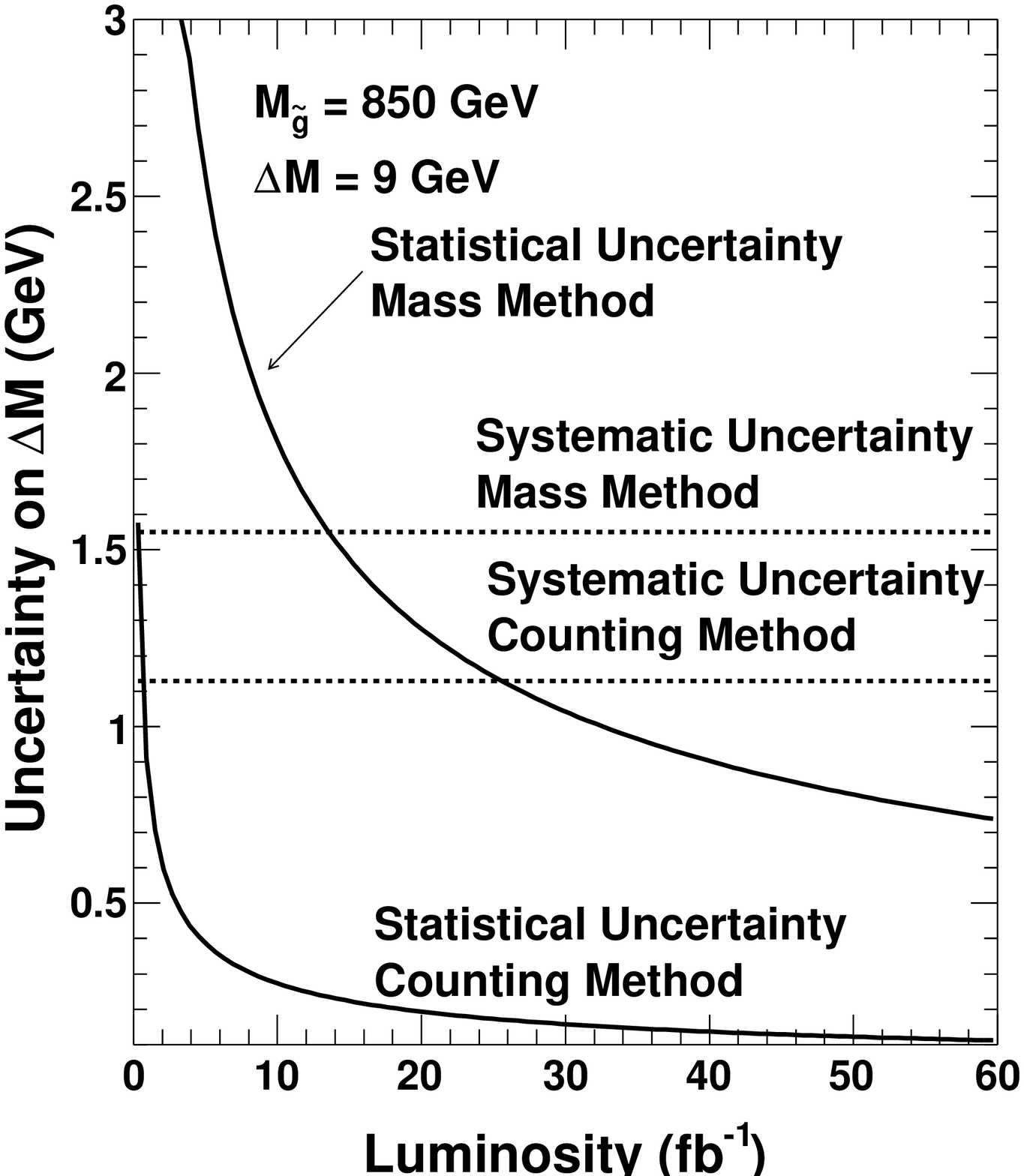}{With the assumption that \mg\ is measured elsewhere to 5\%, we use the \nosls\ and \mtt\ values to make two independent measurements of \dm, which can be combined to produce a more accurate measurement. The top plots show the sources of uncertainty for the counting (left) and the mass (right) methods.  In both cases, the systematic uncertainty dominates the measurement. The bottom plots show combined results as functions of \dm\ (left) and luminosity (right).  We note that both methods are systematics limited by $\mathcal{L}$~=~10~fb$^{-1}$.}{Sigma_All}

\section{Conclusions}
\label{conclusions}

\mysect{conclusions}

We have studied the prospects of simultaneously measuring \dm\ and \mg\ at the LHC in the co-annihilation region using the inclusive 3$\tau+ \mathrm{jet}+\mett$ final state for an mSUGRA scenario. Because this channel is nearly background free after selection cuts, for \dm\ $>$ 5 GeV we are able to form two separate observables: the number of signal events (a \ntwo\ cross-section measurement equivalent) and \mtt.  Both vary with \dm\ and \mg\ allowing the simultaneous measurement of both.  This is particularly important if we are in the co-annihilation region as it is unclear if there will be a high quality \mg\ measurement available since current methods for measuring \mg\ will likely have to be altered to accommodate low \Et\ $\tau$'s in the final state.  We find that with 30~fb$^{-1}$  we can measure \dm\ to 15\% and \mg\ to 6\% for the example point of \dm=9~GeV and \mg=850~GeV.  While our sensitivity to measuring \dm\ at the LHC is not as good as that expected at the ILC, it is quite comparable and should be available much sooner.  Further, a 15\% measurement of \dm\ would generally be sufficient to determine if the signal is consistent with co-annihilation, and therefore, with the \none\ being the dark matter particle. Since no gluino mass measurement is possible at the ILC unless a very high energy option is available, this may be the only correct measurement of \mg.  We also note that we have made no attempt to optimize these results, indicating that with actual data from the detector, we will likely be able to optimize our cuts, leading to a more precise measurement or a lower luminosity needed for the same sensitivity.  We have confirmed that efficient $\tau$ identification down to an \Et\ of 20~GeV is crucial for this analysis as in the 2$\tau$ analysis \cite{two_tau}.  Further, we expect that this analysis and the 2$\tau$ analysis could be used to complement each other in the establishment of a co-annihilation signal at the LHC, and perhaps be combined to produce a more accurate  measurement.  Finally, an analysis of this type can by applied to other SUGRA models provided they have a co-annihilation region and do not suppress the production of the \ntwo\ particles.


\section{Acknowledgments}

This work was supported in part by the Texas A\&M Graduate Merit Fellowship program, DOE grant DE-FG02-95ER40917, and NSF grant DMS 0216275.



\end{document}